# Time-Domain Excitation of Finite-Lifetime Resonances and Their Exceptional Points

Asaf Farhi[1], Dror Hershkovitz[1], Andrea Alù[2,3,4], and Haim Suchowski[1]
*1 School of Physics and Astronomy, Faculty of Exact Sciences, Tel Aviv University, Tel Aviv 69978, Israel*
*2 Photonics Initiative, Advanced Science Research Center, City University of New York, New York, New York 10031, USA*
*3 Department of Electrical Engineering, City College of The City University of New York, New York, New York 10031, and*
*4 Physics Program, Graduate Center, City University of New York, New York, New York 10016, USA*

Resonances associated with complex-frequency poles are ubiquitous across physics and can arise in any open system, ranging from subwavelength particles and cavities to biological structures. When two such resonances coalesce, they form exceptional points (EPs), non-Hermitian singularities known to produce unusual spectral and dynamical behavior. However, the dynamics of the response of such resonances and exceptional points to complex frequency drive remained largely unexplored. Here, we experimentally observe the temporal response of complex-frequency resonances and theoretically study this for exceptional points. We unveil a universal transient phenomenon of open cavities driven at complex frequencies: the system's initial response grows linearly, with enhanced growth at exceptional points (EPs), even though the system is passive and the excitation decays. Closed-form theory for general resonators, extended to higher-order modes, predicts efficient power transfer with $t$ and $t^2$ scaling for complex single poles and exceptional points (EPs), respectively, at all times. We demonstrate these effects in subwavelength optical scatterers and experimentally in an electrical circuit analogue, with excellent agreement, and explore configurations that capture EP-enhanced growth.

Passive open resonators are ubiquitous in both engineered and naturally occurring systems across a wide range of physical domains, including photonics, condensed matter, mechanics, acoustics, thermal physics, and matter waves. Examples of such resonators include cavities, metallic or dielectric subwavelength particles, electric circuits, 2D materials, molecules, ice grains, and droplets, to name a few [1–12]. These resonators inherently possess energy loss mechanisms and therefore are characterized by self-oscillations at complex-valued frequencies. Traditionally they are excited with real-valued frequencies, which are detuned from the true resonant modes of the system. Recently, there has been significant interest in exciting these resonances with more sophisticated waveforms, tailored in time to oscillate at complex-valued frequencies that can engage effectively the poles of interest. These excitation schemes reveal intriguing phenomena, such as overcoming loss in superlensing, enhanced propagation distance of phonon polaritons, and surpassing the scattering limits of passive objects [13–17].

Active resonators can exhibit real frequency poles when material gain precisely balances material and radiative loss, as in a laser at threshold. Significant progress has been made in understanding the temporal dynamics of these systems, and in particular it has been suggested that, under resonant real-valued frequency excitation, their response enhances the order of the input envelope, up to the gain-saturation time [18, 19]. Excitation of larger-than-wavelength passive open resonators at a real frequency (equal to the real part of the complex resonance frequency) has also been recently investigated, and it was shown that their temporal behavior is close to, but distinct, from the one of active resonators at the lasing threshold [20]. More recently, active coupled resonators tuned to a real-frequency exceptional point (EP), where both the eigenstates and resonance poles coalesce [21–24], were predicted to exhibit more exotic temporal dynamics under real-frequency excitations [18].

While experimentally modulating coherent signals is relatively straightforward in the microwave domain, in optics it requires specialized techniques. Such techniques are often discussed in the context of coherent control, in which waveforms are shaped to steer resonant systems [25–30]. Optical waveform tailoring has been key in nuclear magnetic resonance, ultrafast optics, spectroscopy, and quantum dynamics. Typically implemented through spectral pulse shaping, it allows precise control over the amplitude and phase of the excitation field in the frequency domain, which directly governs the system temporal response via Fourier correspondence. Despite challenges in direct time-domain modulation, spectral shaping has enabled transformative applications ranging from selective excitation in molecules to quantum gate operations [31–34]. Recent works have also experimentally demonstrated complex-frequency drive in optics via a superposition of harmonic waves with Lorentzian weighting in the frequency domain [15].

However, time-domain studies to date have primarily focused on real-frequency excitations, and resonant excitations of complex-frequency EPs have remained unexplored, hindering our ability to fully capture the resonant behavior associated with complex-frequency poles and obscuring the dynamical signatures of complex-frequency resonances and EPs. Moreover, the response time of resonators is constrained by the cavity roundtrip time, setting a fundamental limit on the achievable temporal resolution [18–20, 35, 36].

Here, we investigate the excitation of passive resonators with a complex-frequency drive in the time domain. We focus on subwavelength structures composed of metals or dielectric materials, which are widely used in photonics and feature short roundtrip times, thereby enabling substantially enhanced temporal resolution. In analogy, we also study electric circuits that exhibit a similar complex-frequency response but at lower frequencies. We show that, for an incoming waveform with temporal profile $\theta(t)t^m e^{-i\omega_r t-\Gamma t}$, where $\theta(t)$ is the Heaviside function, the response of systems supporting a complex-frequency pole is approximately $\propto t^{m+1}e^{-i\omega_r t-\Gamma t}$, effectively increasing the order of the input envelope. Interestingly, for $t \ll \Gamma$ passive resonators behave similarly to active resonators tuned to the lasing threshold, with approximately a $te^{-i\omega_r t}$ in the response. We also generalize these results to systems exhibiting complex-frequency EPs, where two eigenstates and their corresponding poles become degenerate. Our general result, applicable to passive resonators of any Q-factor, implies counterintuitively that the first-order transient response of the resonant system increases temporally when exciting them at complex frequency with an exponentially decaying amplitude. We analytically derive these results for subwavelength particles, and experimentally demonstrate them in electric circuits, in excellent agreement with theory. We obtain the spatiotemporal dependence of these incoming and outgoing waveforms in the Supplementary Material (SM).

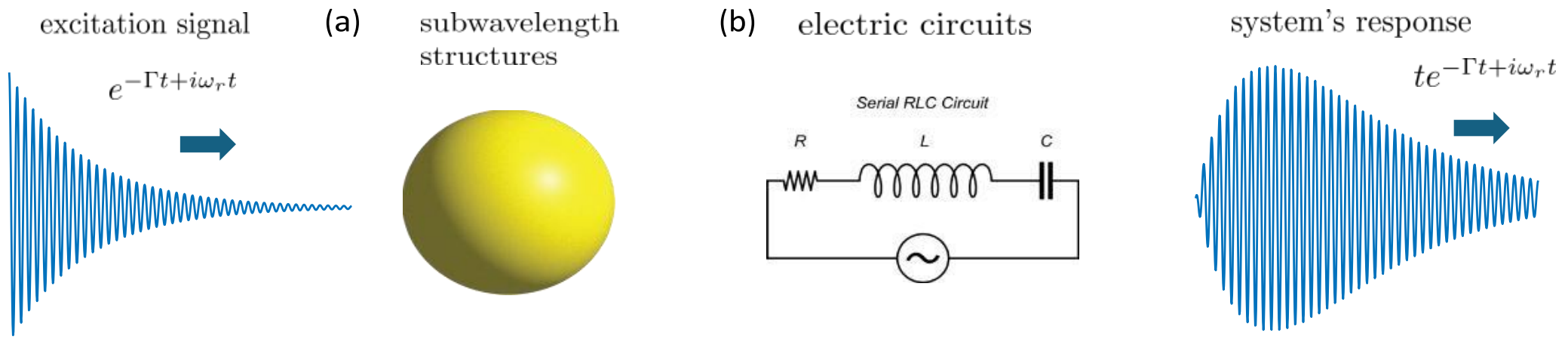

FIG. 1. Passive resonators (without gain) and their temporal response to complex-frequency excitations at resonance. (a) Subwavelength structures, typically composed of plasmon-polariton or phonon-polariton supporting materials. (b) Electric circuits. These resonators are characterized by a complex resonance frequency $\omega_1 = \omega_r + i\Gamma$. As we show, when exciting such resonators with an input wave of the form is $t^m e^{-i\omega_r t-\Gamma t}$, their response is $t^{m+1}e^{-i\omega_r t-\Gamma t}$, which has an approximately $t^{m+1}$ rising-wave envelope when $t \ll 1/\Gamma$. For the waveform presented in the figure, even though both the drive and the mode of the system decay in time exponentially, the initial response increases linearly. A complex-frequency EP further increases the response order.

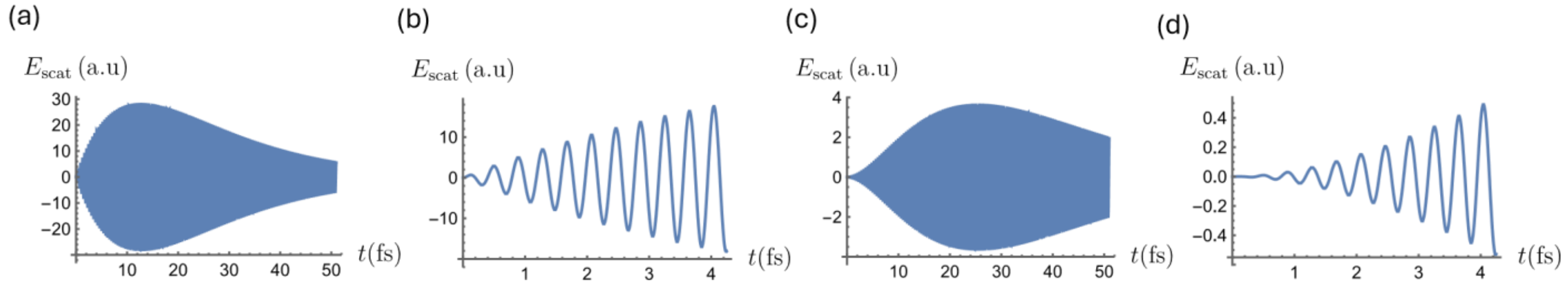

FIG. 2. Response of a subwavelength silver particle with $\omega_p = 1.38 \cdot 10^{16}$ (Hz), $\Gamma = 7.85 \cdot 10^{13}$ (Hz) to complex-frequency resonant excitations. $E_{\rm scat}$ in response to the incoming field $E_{\rm inc} = e^{-i\omega_r t-\Gamma t}$ for the pulse duration (a) and for $t < 1/3\Gamma$ with an approximately linear rise in the field envelope (b). $E_{\rm scat}$ in response to $E_{\rm inc} = \frac{\omega_p}{1000} te^{-i\omega_r t-\Gamma t}$ for the pulse duration (c) and for $t < 1/(3\Gamma)$ with an approximately quadratic rise in the field envelope (d).

Importantly, we find that complex frequency excitations are particularly efficient at engaging the resonance compared with their real-frequency counterpart and exhibit enhanced power transfer. Improving the excitation efficiency of resonances can enhance the performance of various devices used in biomedical electrostimulation, wireless power transfer, RF systems, and miniaturized photonics and electronics [37, 38].

We start by analyzing the response of two types of passive resonators: a subwavelength scatterer and electric circuits with lumped elements. Metallic subwavelength particles have been shown to exhibit resonances in the visible with Q factors in the order of 100. Recent works have also demonstrated phonon-based subwavelength resonant scatterers in the mid-infrared with Q factors in the range of 250-480 [4, 5, 39]. Such subwavelength resonators have a short roundtrip time compared to larger-than-wavelength resonators, providing them with a fast response. For generality, we also derive the non-Hermitian single-mode dynamics in the SM.

We write the scattered field for such a subwavelength structure in the quasistatic regime as [1, 3]

$$E_{\mathrm{sc}} \propto \sum_l \frac{s_l \nabla\phi_l(\boldsymbol{r}) V_{\phi_l E_{\mathrm{inc}}}}{s-s_l}, V_{\phi_l E_{\mathrm{inc}}} = \int \nabla\phi_l(\boldsymbol{r}')\cdot E_{\mathrm{inc}}(\boldsymbol{r}')d\boldsymbol{r}'$$
$$s = \frac{1}{1-\epsilon(\omega)}, s_l = \frac{1}{1-\epsilon_l},\ \epsilon(\omega) = 1-\frac{\omega_p^2}{\omega^2+i\Gamma\omega},$$
$$E_{\mathrm{scat}} \propto \sum_l \frac{\omega_p^2 s_l \nabla\phi_l(\boldsymbol{r}) V_{\phi_l E_{\mathrm{inc}}}}{\omega^2+i\Gamma\omega-s_l\omega_p^2},\ \epsilon_{1l} = -\frac{l+1}{l},$$
$$\omega_{1,2} = i\Gamma \pm \sqrt{4s_{l=1}\,\omega_p^2-\Gamma^2}\ , \quad (1)$$

where $\omega$ is the angular frequency, $\omega_{1,2}$ are the eigenfrequencies, $s_l$ or $\epsilon_l$ is an eigenvalue, $\phi_l$ is an eigenfunction of the source-free Laplace's equation, the integral is performed over the particle volume, and we assume a metallic inclusion characterized by a loss parameter $\Gamma$ and plasma frequency $\omega_p$. Near a resonance, one complex pole dominates the system behavior. Assuming that the dipole mode is dominant, we consider the corresponding eigenvalue $s_{l=1} = 1/3$. We analytically calculate the scattered field in response to a resonant complex-frequency excitation $E_{\mathrm{inc}} = e^{-\Gamma t - i\omega_r t}$ by inverse Fourier transforming $E_{\mathrm{scat}}(\boldsymbol{r},\omega)$ and obtain

$$E_{\mathrm{scat}} = -\frac{\omega_p^2\theta(t)\nabla\phi_1(\boldsymbol{r})V_{\phi_1 E_{\mathrm{inc}}}(-2it\omega_r+e^{2i\omega_r t}-1)e^{-(\Gamma+i\omega_r)t}}{3\cdot 4\omega_r^2} \quad (2).$$

Notably, when $t \gg 1/\omega_r$ the first term dominates anAd we get $E_{\mathrm{scat}}(t) \propto te^{-(\Gamma+i\omega_r)t}$. When also $t \ll 1/\Gamma$ we have approximately an oscillating output with a linearly rising envelope $E_{\mathrm{scat}}(t) \propto te^{-i\omega_r t}$. In general, our analytical model shows that the scattered field has an increase in the order of $t$. For input field $t^m e^{-\Gamma t - i\omega_r t}$, we approximately obtain a scattered field $E_{\mathrm{scat}}(t) \propto t^{m+1} e^{-(\Gamma+i\omega_r)t}$. Such behavior is also expected for subwavelength particles supporting phonon-polariton resonances [4, 5, 39, 40].

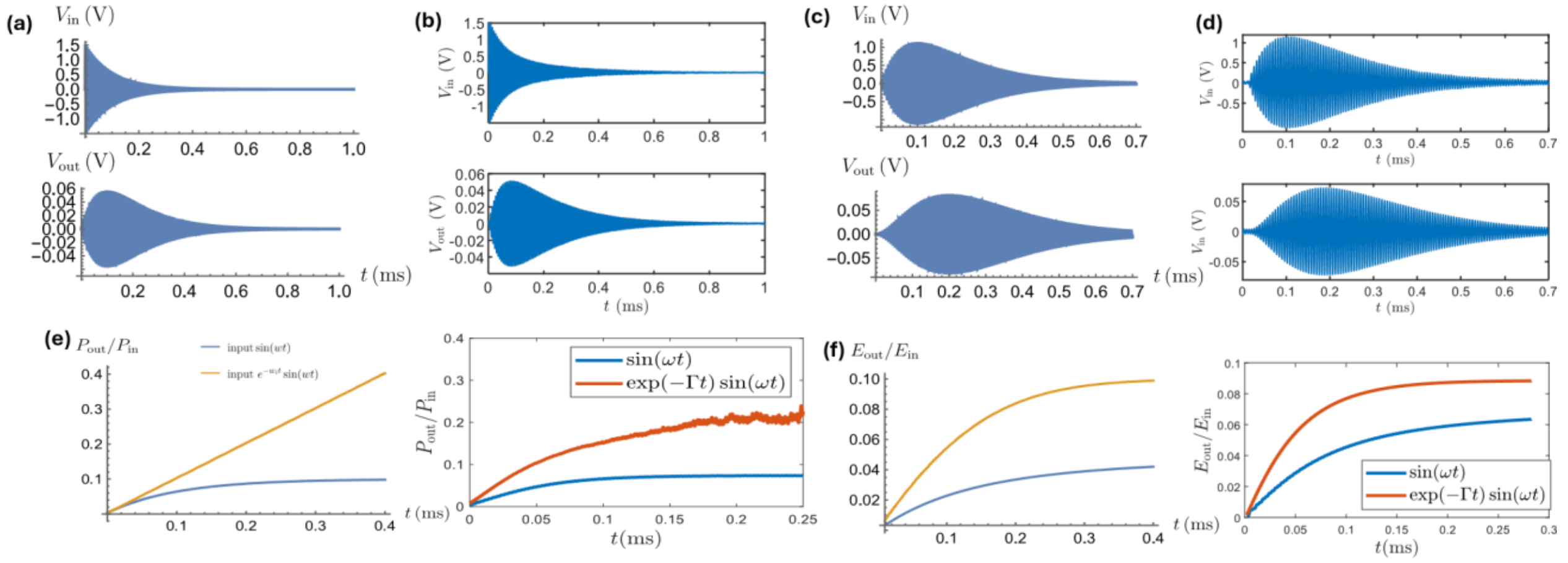

FIG. 3. Theoretical and experimental results for complex-frequency excitations of an electric RLC circuit with $Q = 100, \omega_r = 2\pi\cdot 164100$ (Hz), $\Gamma = 2\pi\cdot 1591.55$ (Hz). The results are similar to those derived for subwavelength particles, but manifest at MHz frequencies. $V_{\mathrm{in}} \propto e^{-i\omega_r t-\Gamma t}$ and $V_{\mathrm{out}} \propto te^{-i\omega_r t-\Gamma t}$ in theory (a) and experiment (b) with excellent agreement. $V_{\mathrm{in}} \propto te^{-i\omega_r t-\Gamma t}$ and $V_{\mathrm{out}} \propto t^2 e^{i\omega_r t-\Gamma t}$ in theory (c) and experiment (d) with very good agreement. For $V_{\mathrm{in}} \propto e^{-i\omega_r t-\Gamma t}$ $P_{\mathrm{out}}/P_{\mathrm{in}}$ (e) and $E_{\mathrm{out}}/E_{\mathrm{in}}$ (f) in theory and experiment, where $P_{\mathrm{out}}$ was measured on the resistor. The values of the electric circuit components are: $R = R_{\mathrm{internal}} + R_{\mathrm{resistor}} = 2 + 16 = 18\Omega, L = 1\mathrm{mH}$, and $C = 0.94\mathrm{nF}$.

Similarly, a lumped-element series RLC circuit has the same denominator form

$$I = \frac{V}{Z} = \frac{V(\omega)j\omega C}{-\omega^2 LC + iRC\omega + 1},$$

$$\omega_{1,2} = i\frac{R}{2L} \pm \sqrt{-\left(\frac{R}{2L}\right)^2 + \frac{1}{LC}}, \quad (3)$$

where $I, V, Z$ are the current, voltage, and total impedance, and $R, L, C$ are the resistance, inductance and capacitance, respectively. We derive the temporal response of an electric circuit for the input voltage $V_{\text{in}}(t) = \theta(t)e^{-i\omega_r t - \Gamma t}$. By inverse Fourier transforming $I(\omega)$ we get

$$I(t) = \frac{C\omega_p^2\theta(t)[-\omega_1^* i\omega_r t e^{-i\omega_r t} + \omega_1 \sin(\omega_r t)]e^{-\Gamma t}}{2\omega_r^2}, \quad (4)$$

where $\omega_{1,2} = \omega_r \pm i\Gamma$. Similarly to the subwavelength particle, after a cycle the second term starts to become negligible, and we get $I(t) \propto te^{-i\omega_r t - \Gamma t}$. Importantly, from this we obtain for the power efficiency $\frac{P_{\text{out}}}{P_{\text{in}}} = \frac{IR}{V_{\text{in}}} \propto t$, for $t > 1/\omega_r$ even though the system is passive. A similar conclusion follows for the high-order complex-frequency drives, which peak at $t_p = \frac{m}{\Gamma}$. In contrast, our calculation for the continuous-wave (cw) excitation shows that at large $t, V_{\text{out}} \propto e^{-i\omega_r t}$ and therefore $\frac{P_{\text{out}}}{P_{\text{in}}}$ is constant. The above results could be understood from the vanishing of the impedance and $|Z| > 0$ for the complex and real-frequency drives, respectively, with the implications for $I$ at large $t$. One can also approximate the excitation efficiency for the inductor and capacitor as $\frac{L\omega_r I}{V}, \frac{I}{C\omega_r V}$ respectively, where here the power is reactive. We thus expect our comparison of the power efficiency on the resistor to also apply to such comparisons for the inductor and capacitor.

Note that this is a quasistatic or lumped element approximation and in practice, since the constructive interference, on which the resonance effect is based, starts after a roundtrip, the resonator size limits the response speed in both cases. While this effect is neglected in the modeling of subwavelength structures, it is expected to discretize the response. Clearly, in these two cases the response speed is much faster than for the previously considered larger-than-wavelength resonators. Crucially, for the complex-resonance EP, excited by the complex-frequency excitation, we obtained $I(t) \propto t^2 e^{-i\omega_r t - \Gamma t}$, with the order of $t$ increasing by the EP order (SM). The power efficiency in this case has a $t^2$ scaling for all $t$, including for high-order drives.

To demonstrate the response of the subwavelength particle to resonant complex-frequency excitations, we consider a silver particle with $\omega_p = 1.38 \cdot 10^{16}$ (Hz), $\Gamma = 7.85 \cdot 10^{13}$ (Hz), and $R = 40$ (nm), where $R$ is the particle radius. In Fig. 2 (a) and (b) we plot the analytically calculated response to an incoming field of the form $E_{\text{inc}} = \theta(t)e^{-\Gamma t + i\omega_r t}$ for the pulse duration and for $t < 1/(3\Gamma)$, respectively. In Fig. 2 (c) and (d) we plot the response to $E_{\text{inc}} = \theta(t)\frac{\omega_p}{1000}te^{-\Gamma t - i\omega_r t}$ for the pulse duration and for $t < 1/(3\Gamma)$, respectively. Importantly, in the region $t < 1/(3\Gamma)$ the responses shown in Fig. 2 (b) and (d) are similar to the real-frequency resonance case, with linear and quadratic rises of the scattered-field envelopes, respectively.

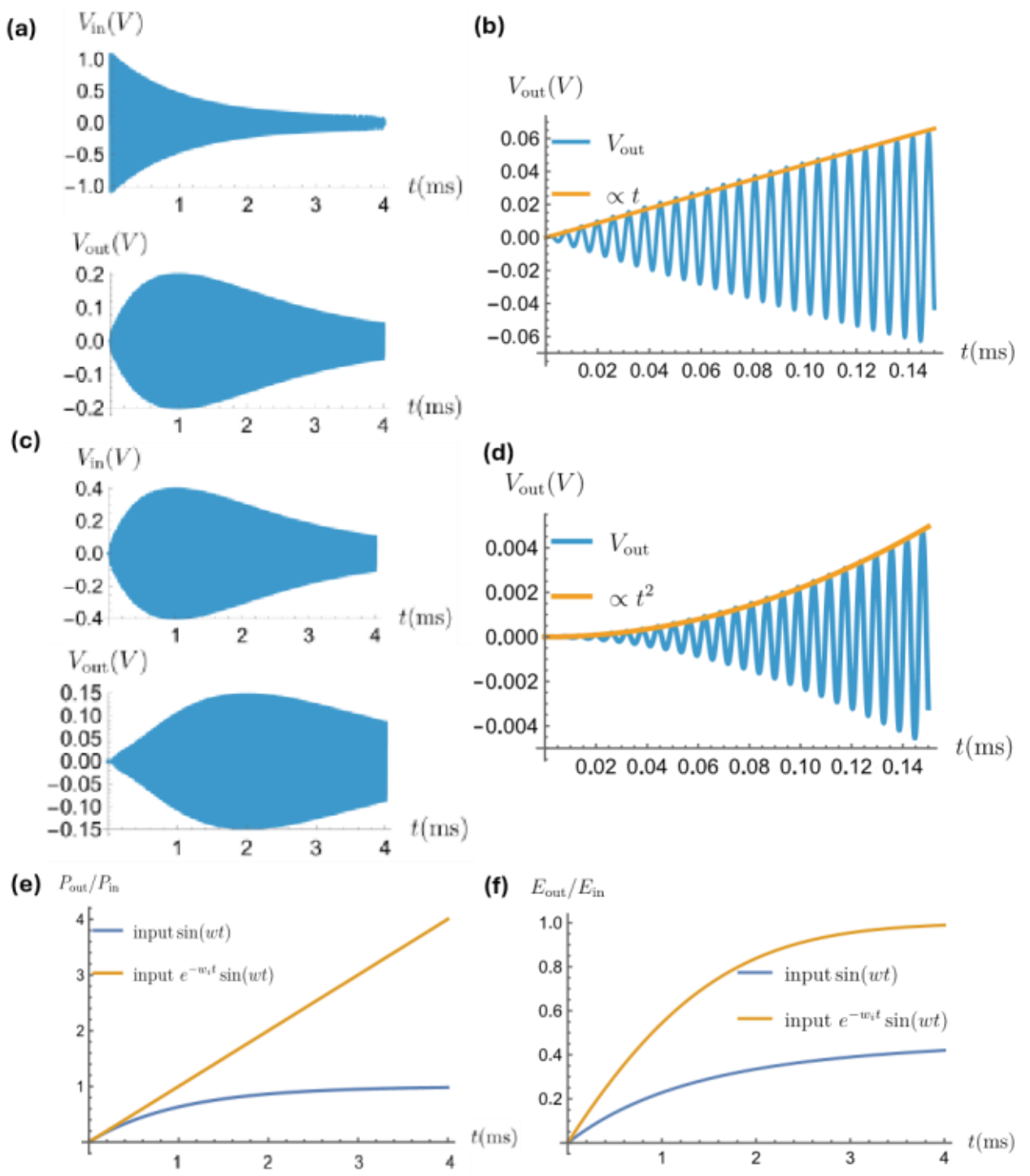

FIG. 4. Analytical results for complex-frequency excitations of an electric RLC circuit with $Q = 1000$ and $\omega_r = 2\pi \cdot 164100$ (Hz), $\Gamma = 2\pi \cdot 159.15$ (Hz). For $V_{\text{in}} \propto e^{-i\omega_r t - \Gamma t}$, $V_{\text{out}} \propto te^{-i\omega_r t - \Gamma t}$ for the pulse duration (a) and for $t < 0.15/\Gamma$ with excellent match to a $\propto t$ envelope (b). For $V_{\text{in}} \propto te^{-i\omega_r t - \Gamma t}$, $V_{\text{out}} \propto t^2 e^{i\omega_r t - \Gamma t}$ for the pulse duration (c) and for $t < 0.15/\Gamma$ with excellent match to a $\propto t^2$ envelope (d). $P_{\text{out}}/P_{\text{in}}$ (e) and $E_{\text{out}}/E_{\text{in}}$ (f) for $V_{\text{in}} \propto e^{-i\omega_r t - \Gamma t}$, where $P_{\text{out}}$ and $E_{\text{out}}$ are calculated on the resistor, demonstrating superior power efficiency of complex-frequency excitations independently of the Q factor.

To observe this behavior, we analytically calculate and perform experiments on a series RLC circuit with Q=100. We consider the input voltages $e^{-i\omega_r t - \Gamma t}$, $te^{-i\omega_r t - \Gamma t}$, and $e^{-i\omega_r t}$, where $\omega = \text{Re}(\omega_1)$ and $\Gamma = \text{Im}(\omega_1)$, and calculated and measured the output voltage on the resistor. The RLC values were $R = R_{\text{internal}} + R_{\text{resistor}} = 2 + 16 = 18\Omega, L = 1\text{mH}$, and $C = 0.94\text{nF}$. We also calculated the ratios of the output power measured on the resistor to the input power, and output energy to input energy for the first excitation and compare them to the results for the standard cw excitation. We then performed the same calculations for an RLC circuit with Q=1000 without performing experiments as an implementation of electric circuits with high Q factors is rather challenging. In Fig. 3 we show the calculations and experimental results for the series RLC circuit with Q = 100. In Fig. 3 (a) and (b) we present the theoretical and experimental $V_{\text{in}}$ and $V_{\text{R}}$, respectively, for $V_{in} = \sin(\omega_r t)e^{-\Gamma t}$, showing excellent agreement. It can be seen that $V_{\text{out}} \approx t\sin(\omega_r t)e^{-\Gamma t}$ with an envelope of $t$ for $t \ll 1/\Gamma$, which is around 5 cycles ($te^{-\Gamma t} \approx t$ in expansion). Here, initially, the measured input voltage decayed faster than the applied input voltage due to impedance mismatch and to alleviate it, we utilized an applied input voltage with a slower decay rate and obtained the required measured input voltage (note that this impedance matching is transient as opposed to the standard impedance matching in frequency domain). In Fig. 3 (c) and (d) we show the theoretical and experimental results for $V_{\text{in}} = t\sin(\omega_r t)\,e^{-\Gamma t}$ with very good agreement. As we predicted $V_{\text{out}} \approx t^2\sin(\omega_r t)\,e^{-\Gamma t}/2$ with a $t^2/2$ envelope for $t \ll 1/\Gamma$. Fig. 3 (e) and (f) show the theoretical and experimental results of $P_{\text{out}}/P_{\text{in}}$ and $E_{\text{out}}/E_{\text{in}}$, respectively, where $E_{\text{out/in}} = \int_0^t P_{\text{out/in}}(t')dt'$, for $V_{\text{in}} = \sin(\omega_r t)e^{-\Gamma t}$ and cw excitation, with qualitative agreement. Remarkably, the complex-frequency excitation has superior performance in both cases. Note that even if the cw excitation is switched off to improve excitation efficiency, it still does not offer the same level of control as complex-frequency excitation, which is especially important in multi-channel systems [41].

In Fig. 4 we present our calculations for the RLC circuit with Q=1000. Fig 4 (a) and (b) show the output voltage in response to $V_{\text{in}} = \sin(\omega_r t)e^{-\Gamma t}$ for the pulse duration and for $t \ll 1/\Gamma$, respectively. Here, due to the higher Q factor, the envelope of $V_{\text{out}}$ scales as $t$ for more cycles compared with the previous Q=100 circuit. Fig 4 (c) and (d) present the response to $V_{\text{in}} = t\sin(\omega_r t)e^{-\Gamma t}$ for the pulse duration and for $t \ll 1/\Gamma$, respectively. Here, too, the envelope of $V_{\text{out}}$ is proportional to $t^2$ for more cycles. Fig. 4 (e) and (f) also show superior performance for $P_{\text{out}}/P_{\text{in}}$ and $E_{\text{out}}/E_{\text{in}}$, which underscores the independency of these results on the Q factor.

Finally, we study the dynamical behavior of systems tuned to complex-frequency EPs, where two modes and eigenfrequencies coalesce. In general, degeneracies may arise from symmetry or emerge accidentally in non-Hermitian systems, with the latter potentially leading to EPs. In particular, we show that a two-slab setup can be tuned to a complex-frequency EP and, despite being passive, exhibits enhanced growth in response to an exponentially-decaying wave. Note that previous works have mostly focused on real-frequency EPs [22, 36, 42, 43], and here we focus on the unique behavior of purely passive systems tuned to complex-frequency EPs. The total reflectivity of this structure is given by

$$r = \frac{r_1(e^{2ikn_2l_2}r_2 + 1) + e^{2ikn_1l_1}(r_2 + e^{2ikn_2l_2})}{1 + e^{2ikn_2l_2}r_2 + e^{2ikn_1l_1}(r_2 + e^{2ikn_2l_2})r_1} \approx \frac{r_1(e^{2ikn_2l_2}r_2 + 1) + e^{2ikn_1l_1}(r_2 + e^{2ikn_2l_2})}{2i\frac{\omega-\omega_1}{c}n_2l_2r_2 + 2i\frac{\omega-\omega_1}{c}n_1l_1\left(r_2 + 2i\frac{\omega-\omega_1}{c}n_2l_2\right)r_1}$$

where $k$ is the wavevector, $r_1, r_2$ are the interface reflectivitties, $l_1, l_2$ are the slab lengths, and $n_1, n_2$ are the refractive indices of the slabs. In the last transition we expanded the denominator to first order in $\omega$ close to a pole $\omega_1$ to underscore the quadratic behavior using the resonance condition. When the drive frequency $\omega_d \approx \omega_1$, which can be realized via $E_{\text{inc}} \propto e^{-\Gamma t - i\omega t}$ at large $t$, and $\omega_1 = \omega_{\text{EP}}$, $r$ can be approximated by a double complex-pole structure that yields via an inverse Fourier transform $E_{\text{scat}} \propto t^2 e^{-\Gamma t - i\omega t}$. In practice, the roundtrip introduces a discretization effect to the response [44]. In the SM we studied both larger-than-wavelength slabs and slabs on the order of the wavelength with a high Q factor, and focus here on the latter for enhanced temporal resolution. In Fig. 5 (a) we present a realization of the setup with Bragg mirrors to increase the reflectivity and enable relatively continuous response as explained in the SM. Fig. 5 (b) and (c) demonstrate the signature of the EP for the one we calculated, with quadratic dependency of the response as a function of $\omega$ and coalescence of eigenfrequencies, respectively (we use natural units with $c = 1$). In Fig. 5 (d) we show the scattered field as a function of time for a resonant exponentially-decaying drive $E_{\text{inc}} \propto e^{-i\omega_r t - \Gamma t}$, which we calculated numerically by decomposition to spectral regions [18]. Remarkably, even though the incident field decays and the system is passive, the response increases quadratically. We derive the general result for complex-frequency EP using a coupled-mode approach in the SM. In SM we present a passive electric circuit tuned to complex-frequency EPs, underscoring the universality of this effect.

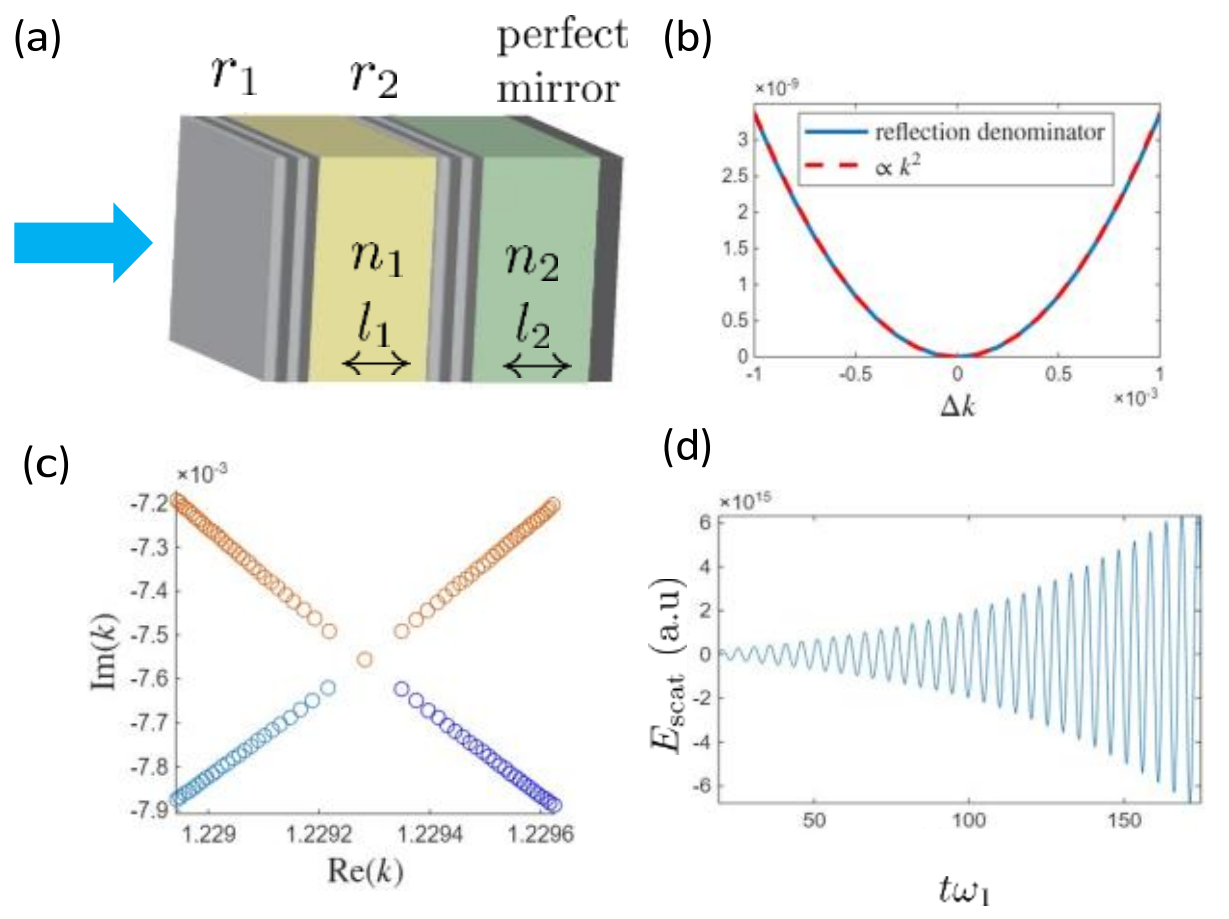

FIG. 5. Two-slab one-sided system with Bragg mirrors exhibiting complex-frequency exceptional points, enabling tunability of the Q factor. (a) The two-slab setup with Bragg mirrors is terminated by a perfect mirror on the right. (b) Quadratic behavior of the denominator of the total reflection coefficient as a function of frequency $r(\omega) \propto 1/(\omega-\omega_1)^2$ near the complex-frequency EP. (c) Convalescence of two eigenfrequencies at the complex-frequency EP as $n_2$ is varied. (d) The temporal response to an exponentially-decaying resonant incident field, showing a quadratic increase of the scattered field even though this system is passive.

In conclusion, we analyzed the temporal response to resonant complex-frequency excitations of a class of resonant physical systems that support complex-frequency poles. We studied two types of passive resonators: subwavelength particles and electric circuits, and experimentally validated our theory with the latter. We showed that excitations of the form $e^{i\omega_r t-\Gamma t}$ approximately result in an output $te^{-i\omega_r t-\Gamma t}$. For times much shorter than $1/\Gamma$, this resembles the functioning of an active real-frequency resonator with the response $te^{-i\omega_r t}$ . We generalized these results for complex-frequency EPs, which further increase the order of the input envelope. Finally, we showed that complex-frequency excitations provide superior power efficiency, potentially enabling enhanced selectivity [41], which can be utilized for various applications including biomedical electrostimulation (e.g., tumor treating fields), wireless power transfer, RF systems, and miniaturized photonics and electronics [37, 38]. Very recently, it was shown theoretically that atoms and molecules behave similarly to this class of passive resonators but with the Rabi frequency playing an important role and can process waves with attosecond resolution [44]. In this work, the response of a high-Q larger-than-wavelength resonator was also shown to exhibit a discretization effect arising from the roundtrip, which modified the response $te^{-i\omega_r t-\Gamma t}$ [44]. Moreover, the influence of nearby poles was analyzed in atomic systems, potentially leading to beating at long times [44]. Future directions include selective excitation of a single system pole in multi-channel systems via a complex-frequency resonant drive.

**ACKNOWLEDGEMENT**

E. Shaked is acknowledged for constructing the electric circuits. Y. Hadad is acknowledged for the useful comments.